# Tunable intersublattice exchange coupling drives magnetic evolution in Mn$_{3+x}$Ga$_{1-x}$C (0 ≤ $x$ ≤ 0.60)

Dong-Hui Xu, Cong-Mian Zhen, Deng-Lu Hou, Li Ma, De-Wei Zhao, Guo-ke Li*

*Hebei Advanced Thin Films Laboratory, College of Physics, Hebei Normal University, Shijiazhuang, 050024, China.*

**Abstract:** We investigate the magnetic and transport evolution in Mn$_{3+x}$Ga$_{1-x}$C (0 ≤ $x$ ≤ 0.60), where Mn substitution at corner Ga sites induces lattice contraction and suppresses the antiferromagnetic order of Mn$_3$GaC. As $x$ increases, the magnetic ground state of the system undergoes a sequential transition from an antiferromagnetic state, via a canted ferrimagnetic state, to a robust ferrimagnetic state, accompanied by a surge in the magnetic ordering temperature. Saturation magnetic moments reaches a maximum of 3.63 μ$_B$/f.u. at $x$ = 0.10, whereas the topological Hall resistivity peaks at 1.47 μΩ·cm for $x$ = 0.20 before decreasing with further doping. First-principles calculations demonstrate a ~40° canting of face-centered Mn moments at $x$ = 0.20, signifying spin frustration, and an eventual antiparallel alignment of face-centered and corner-site Mn moments at higher $x$. These results reveal that intersublattice antiferromagnetic coupling governs the magnetic transformation and emergent transport phenomena, thus providing a microscopic foundation for designing high-ordering-temperature antiperovskites.

*Contact author: liguoke@126.com

## 1. Introduction

Antiperovskite Mn$_3$GaC, characterized by strong spin-lattice coupling[1-3], exhibits multifunctionality of magnetovolume, magnetocaloric[4-7], and giant magnetoresistance effects[8]. As depicted in Fig. 1(a), its cubic unit cell features Mn atoms at face-centered sites (Mn-I), Ga at the corner sites, and C at the body-center site. A delicate balance of intrasublattice exchange interactions among Mn-I stabilized an antiferromagnetic ground state with spins aligned along the [111] direction[9]. Upon heating, the system, accompanied by negative thermal expansion, transitions from antiferromagnetic to a canted ferromagnetic state near 150 K, evolves into a ferromagnetic state around 160 K, and enters a paramagnetic state above a Curie temperature ($T_C$) of 250 K[10-13]. A key strategy to tailor the magnetic properties of this system involves chemical tuning via partial substitution of corner site Ga by excess Mn (designated Mn-II), yielding the non-stoichiometric series Mn$_{3+x}$Ga$_{1-x}$C, which displays complex magnetism and a dramatically enhanced magnetic ordering temperature[14-16].

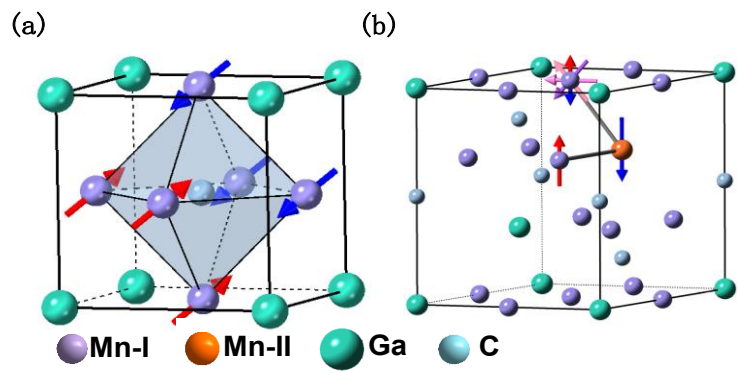

Fig. 1. Schematic illustration of the magnetic configuration in Mn$_{3+x}$Ga$_{1-x}$C. (a) Unit cell of cubic antiperovskite Mn$_3$GaC, featuring alternating antiferromagnetically coupled (111) planes; (b) R-3m unit cell of Mn$_{3+x}$Ga$_{1-x}$C, obtained by projecting the parent antiperovskite structure along the [111]

direction, consisting of Kagome layers of Mn and C interleaved with C sheets; The Mn-II atom substituting Ga cannot align antiparallel to all twelve neighboring Mn-I atoms, leading to spin frustration within the Kagome lattices.

Structurally, stochiometric $Mn_3GaC$ consists of alternating Kagome layers of Mn-I and Ga atoms, interleaved with C sheets along the [111] direction. In $Mn_{3+x}Ga_{1-x}C$, the Mn-II have a 2.75 Å distance with neighboring Mn-I atoms[10]. This proximity, below the 2.90 Å threshold for Mn–Mn exchange coupling, fosters a robust intersublattice antiferromagnetic exchange interaction, as evidenced in the high Néel temperature ($T_N$ > 850 K) for the ferrimagnetic endmember $Mn_4C$[17-21]. As illustrated in Fig. 1(b), the exchange interaction of Mn-I and Mn-II is geometrically and magnetically incompatible with the intrinsic antiferromagnetic order of pristine $Mn_3GaC$. At low $x$, the incorporation of Mn-II moments into the Kagome plane introduces magnetic frustration due to competing interactions, which is closely associated with non-collinear spin textures[2, 22, 23]. With increasing $x$, the Mn-I and Mn-II interaction progressively dominates the magnetic exchange landscape, ultimately stabilizing a long-range ferrimagnetic state reminiscent of $Mn_4C$. Thus, the competition between the native Kagome antiferromagnetism and the emergent Mn-I and Mn-II intersublattice coupling provides a unified framework for understanding the composition-dependent evolution of both magnetic order and transport behavior in $Mn_{3+x}Ga_{1-x}C$. Despite this framework, the microscopic role of Mn-I and Mn-II exchange interactions in mediating the crossover from frustrated antiferromagnetism to ferrimagnetism, and its direct influence on electrical transport, remains poorly understood and warrant systematic investigation.

Motivated by this gap, this work provides a combined theory and experiment investigation of $Mn_{3+x}Ga_{1-x}C$ ($0 \leq x \leq 0.60$). The results demonstrate that increasing $x$ drives a continuous magnetic evolution, from an antiferromagnetic state, via a spin-frustrated state, to a ferrimagnetic state, accompanied by a marked elevation of the magnetic ordering temperature. These findings unambiguously identify that the exchange interaction between Mn-I and Mn-II governs both magnetic architecture and emergent electronic transport properties of this compositionally tunable antiperovskite.

## 2. Experimental

Polycrystalline $Mn_{3+x}Ga_{1-x}C$ ($x$ = 0.00, 0.05, 0.10, 0.20, 0.40, 0.60) samples were synthesized by solid-state reaction. Stoichiometric amounts of high-purity Mn, Ga, and C—supplemented with an additional 5 at.% C to compensate for potential losses—were sealed in evacuated quartz tubes. The mixtures were first annealed at 500 °C for 12 h to promote homogenization, then reground, pelletized, and sintered at 800 °C for 48 h. To ensure phase purity and complete reaction, the pellets were reground, repalletized, and subjected to a second sintering step under identical conditions. The crystal structures were characterized by X-ray diffraction (XRD, PANalytical). Magnetic and electrical transport properties were measured using a Physical Property Measurement System (PPMS-9, Quantum Design).

The magnetic configuration of $Mn_{3+x}Ga_{1-x}C$ was determined via the VASP code within the framework of generalized gradient approximation (GGA). To accommodate the magnetic ordering, the cubic Pm-3m unit cell is symmetry-lowered to a trigonal R-3m unit cell with $\tilde{a} = \tilde{b} = \sqrt{2}a$ and $\tilde{c} = \sqrt{3}a$, where $a$ denotes the XRD-refined cubic lattice parameter. Static calculations were performed on 2×2×1 R-3m supercells, with Mn-I moments constrained to a fixed polar angle $\theta$ within the [0001]-containing plane and Mn-II magnetic moments fixed along [0001] direction. The ground-state magnetic configuration was identified by minimizing the total energy as a function of $\theta$. Brillouin-zone integration was implemented with a Γ-centered 9×5×2 $k$-point mesh, together with a plane-wave cutoff energy of 500 eV and an electronic self-consistency tolerance of $10^{-6}$ eV.

## 3. Results and Discussion

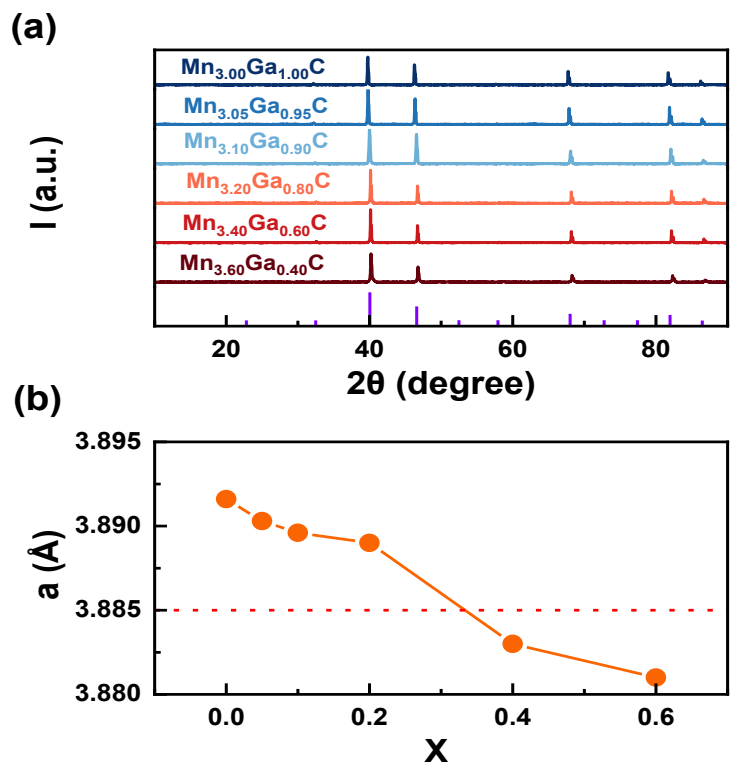

**Fig. 2.** Structural characterization of $Mn_{3+x}Ga_{1-x}C$ samples. (a) XRD patterns of $Mn_{3+x}Ga_{1-x}C$ for $x$ = 0.00, 0.05, 0.10, 0.20, 0.40, and 0.60. (b) Lattice constant $a$, refined from Rietveld analysis, as a function of Mn substitution level $x$.

The XRD patterns of the polycrystalline $Mn_{3+x}Ga_{1-x}C$ ($0 \leq x \leq 0.60$) samples are presented in Fig. 2(a). All observed peaks can be indexed to the antiperovskite structure, with no indication of secondary phase impurities across any of the compositions examined. As the substitution of Ga by Mn atoms increases, a systematic shift of all diffraction peaks toward higher angles is noted, indicating lattice contraction. This observation is further supported by Rietveld refinement of the XRD data, which yielded $\chi^2$ values less than 0.5, indicating high phase purity and excellent crystallinity of the samples. The refined lattice parameter, depicted in Fig. 2(b), records a significant reduction from 3.891 Å at $x = 0.0$ to 3.883 Å at $x = 0.4$, followed by a more gradual decrease to 3.881 Å at $x = 0.6$. This trend demonstrates the lattice contraction effect induced by Mn substitution at the corner Ga sites, thus reflecting the impact of Mn incorporation on the structural properties of $Mn_{3+x}Ga_{1-x}C$.

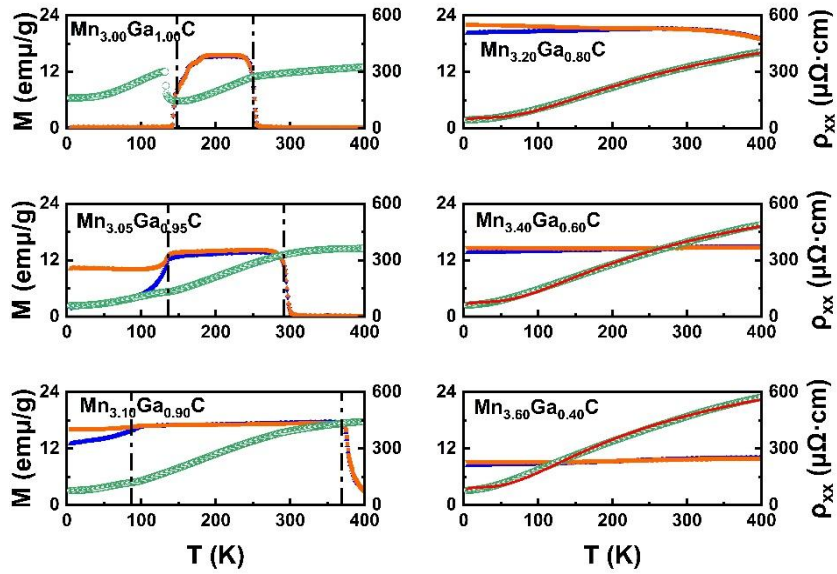

**Fig. 3.** Temperature-dependent magnetic and electrical properties of $Mn_{3+x}Ga_{1-x}C$. Zero-field-cooled (ZFC) and field-cooled (FC) magnetization measured under a 100 Oe applied field (left axis), and electrical resistivity (right axis) are plotted for compositions with (a) $x = 0.00$, (b) $x = 0.05$, (c) $x = 0.10$, (d) = 0.20, (e) $x = 0.40$, and (f) $x = 0.60$.

Figure 3 presents the temperature-dependent zero-field-cooled (ZFC) and field-cooled (FC) magnetization curves of $Mn_{3+x}Ga_{1-x}C$ measured under a 100 Oe magnetic field (left axis), along with the electrical resistivity curves (right axis). For the undoped sample, a transition from an antiferromagnetic to a canted ferromagnetic state occurs at

143 K, followed by a transition to a collinear ferromagnetic state at 178 K and finally a transition to paramagnetic state at 252 K. Correspondingly, the resistivity curve exhibits a minimum near 140 K and a kink around 250 K, consistent with previous reports on $Mn_3GaC$. Mn substitution on the Ga sublattice completely suppresses the antiferromagnetic ground state and replaces the native ferromagnetic order with ferrimagnetic order. As $x$ increases, the canted-ferrimagnetic to ferrimagnetic transition temperature decreases monotonically to 133 K at $x = 0.05$, and further to 93 K at $x = 0.10$, with clear dips in the resistivity curve at both temperatures reflecting strong magnet-transport coupling. Concurrently, the $T_N$ rises steadily to 296 K and 376 K for $x = 0.05$ and 0.10, respectively. For $x \geq 0.20$, the samples maintain ferrimagnetic order across the entire temperature range (with $T_N > 400$ K), and the resistivity obeys the Bloch–Grüneisen law, which is dominated by electron–phonon scattering[24, 25].

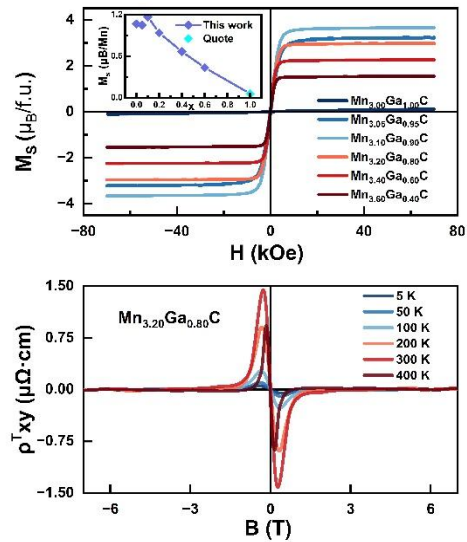

**Fig. 4.** Evolution of magnetism and topological Hall effect in $Mn_{3+x}Ga_{1-x}C$. (a) Magnetic hysteresis loops measured at 5 K across the doping series. The inset plots the saturation magnetic moment ($M_S$)

per Mn atom versus doping level $x$, with the Mn$_3$GaC value obtained in its ferromagnetic state (150 K) included for reference. (b) Field dependence of the topological Hall resistivity ($\rho_{xy}^T$) for Mn$_{3.2}$Ga$_{0.8}$C at selected temperatures.

Figure 4(a) presents hysteresis loops of Mn$_{3+x}$Ga$_{1-x}$C measured at 5 K. In contrast to the negligible magnetization for Mn$_3$GaC, all Mn-doped Mn$_{3+x}$Ga$_{1-x}$C samples exhibit distinct ferromagnetic-like hysteresis. The evolution of the saturation magnetic moment (M$_S$) with doping level $x$ is summarized in the inset of Fig. 4(a). Starting from approximately 1.0 μ$_B$/Mn in Mn$_3$GaC (taking the value at 150 K), m$_S$ increases to a maximum of 1.17 μ$_B$/Mn at 0.10, before decreasing monotonically to 0.42 μ$_B$/Mn at 0.60. The notable reduction in m$_S$ confirms the antiparallel coupling between Mn-I and Mn-II moments, while the non-monotonic evolution suggests that doping progressively modifies the canting angle of the Mn-I sublattice relative to Mn-II. Furthermore, the increase in $x$ enhances the intersublattice antiferromagnetic couplings, which leads to the pronounced rise in the T$_N$ observed in Fig. 3. Collectively, the magnetic results provide compelling evidence that the tunable antiferromagnetic coupling between the Mn-I and Mn-II sublattices governs the evolution of magnetic order in Mn$_{3+x}$Ga$_{1-x}$C.

As mentioned above, incorporating incompatible Mn-II moments into the Kagome plane of Mn$_3$GaC introduces non-collinear spin textures, which generate a real-space Berry curvature and readily give rise to the topological Hall effect, a key transport signature for probing magnetic topology[26-29]. To elucidate the evolution of topological magnetism in this system, we systematically measured the transverse resistivity of Mn$_{3+x}$Ga$_{1-x}$C and isolated the topological Hall resistivity[30, 31], $\rho_{xy}^T$, via multi-field analysis. It is found that the magnitude of $\rho_{xy}^T$ strongly depends on both composition and temperature, directly evidencing strong coupling between magnetic texture and charge transport. As illustrated in Fig. 4(a) for $Mn_{3.2}Ga_{0.8}C$, $\rho_{xy}^T$ increases with temperature, peaks at 1.47 μΩ·cm near 300 K, and then decreases to 0.94 μΩ·cm at 400 K. These finding reveals the stabilization of non-collinear magnetic order in $Mn_{3+x}Ga_{1-x}C$.

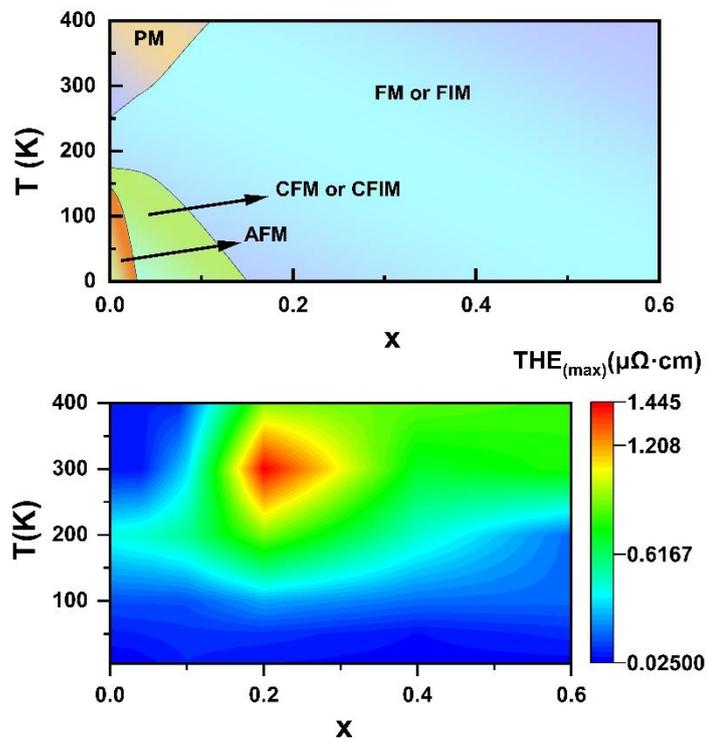

**Fig. 5.** Magnetic and topological Hall effect phase diagrams of $Mn_{3+x}Ga_{1-x}C$. (a) Magnetic phase diagram as a function of $x$ and temperature, constructed from magnetization measurements. (b) Contour map of the topological Hall resistivity, $\rho_{xy}^T$, in the x–T parameter space, highlighting the evolution of non-collinear spin textures.

To establish a direct relationship between magnetic phase evolution and the emergence of topological transport, Figure 5 presents the phase diagrams and contour map of $\rho_{xy}^T$ of $Mn_{3+x}Ga_{1-x}C$ as functions of composition $x$ and temperature $T$. As shown in Fig. 5(a), increasing $x$ drives the magnetic state of $Mn_{3+x}Ga_{1-x}C$ evolving from antiferromagnetic state, through an intermediate spin-frustrated state, toward a ferrimagnetic order[32, 33]. In contrast, Fig. 5(b) reveals that the topological Hall resistivity attains its maximum value near $x = 0.2$ and 300 K, which indicates substantial real-space Berry curvature associated with the non-collinear spin textures[26, 27]. This correlation directly highlights significant spin frustration around $x = 0.20$, which stabilizes non-collinear magnetic orders in $Mn_{3+x}Ga_{1-x}C$. These findings provide critical insights into how the interplay between Mn-I and Mn-II sublattices governs the magnetic configuration, and underscores the importance of Mn substitutions in tailoring magnetic and electronic properties of antiperoverskites.

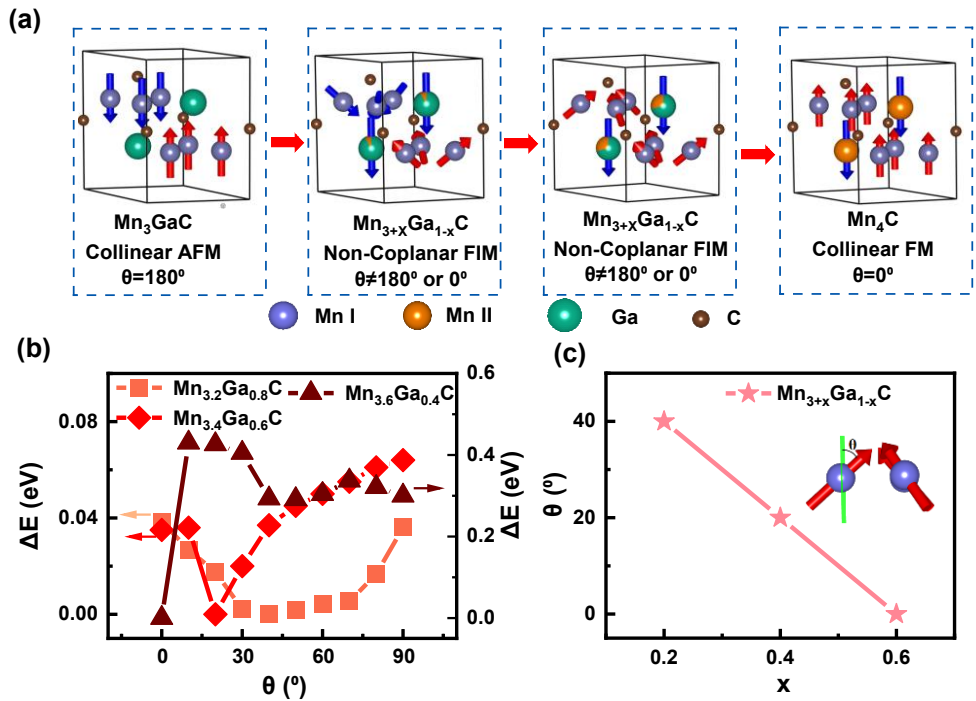

**Fig. 6.** Calculation of atomic spin angle variation in $Mn_{3+x}Ga_{1-x}C$. (a) Schematic diagram of crystal and magnetic structure of $Mn_{3+x}Ga_{1-x}C$ ($0.00 \leq x \leq 1.0$), with Mn I,

Mn II, Ga, and C atoms represented by purple, orange, green, and brown, respectively; (b) Dependence of energy difference ΔE on angle $\theta$ at different Mn doping levels, where ΔE is defined as the difference between the ground state energy at a given $\theta$ and the global minimum energy calculated from the experimental lattice constant; (c) Variation of stable spin angle $\theta$ with x content, where $\theta$ is the angle between the spin direction and the vertically upward direction, as shown in the inset.

From an experimental perspective, we can macroscopically observe that the amplitude of $\rho_{xy}^T$ increases significantly at $x = 0.2$, and it reaches saturation even at low fields. This indicates that the formation energy barrier of the topological spin texture is low at this point, making it easy to be stabilized by an external magnetic field, providing an important basis for the production of non-coplanar magnetic structures. Next, we will further verify this through theoretical calculations. Fig. 6(a) shows the evolution of atomic spin configuration and magnetic order in the $Mn_{3+x}Ga_{1-x}C$ ($0.00 \leq x \leq 1.0$) system as the Mn doping amount x changes. In pure-phase $Mn_3GaC$, the system exhibits collinear antiferromagnetic (CFM) order, with Mn atomic spins aligned in opposite directions at $\theta = 180°$. As the Mn content increases ($x > 0$), the system transitions to a non-coplanar ferrimagnetic (Non-Coplanar FIM) state, where the spins of Mn I and Mn II atoms become tilted, forming a non-coplanar magnetic structure with $\theta \neq 180°$ or $0°$. When x approaches 1.0, the system eventually evolves into a collinear ferromagnetic (CFM) state, where all Mn II atoms align their spins in the same direction, and $\theta = 0°$. This continuous magnetic phase transition process clearly reveals the regulatory effect of Mn doping on the spin interaction and magnetically ordered state of the system. By performing static energy calculations on the magnetic structure at different angles ($\theta$ angle is shown in the inset of Fig. 6(c), where calculations are performed from 0 to 90°), ΔE is defined as the difference between the state energy at a given θ and the global minimum energy value[9]. The calculated energy differences at various angles are shown in Fig. 6(b). For $x$ = 0.0, the minimum energy value occurs at $\theta$ = 180°, corresponding to the collinear. antiferromagnetic ground state. Analysis of the calculation results, as shown in Fig. 6(c), reveals that as x increases to 0.2, the calculated magnetic ground state is a non-coplanar magnetic structure with an angle of 40°, which corroborates the experimentally observed large Top Hall resistivity. As x increases to 0.4 and 0.6, the minimum energy value gradually shifts towards $\theta$ = 0°, corresponding to a collinear subferromagnetic ground state. These results quantitatively confirm the energy-driven mechanism of the magnetic phase transition. The study systematically elucidates that in the $Mn_{3+x}Ga_{1-x}C$ system, Mn doping achieves a continuous magnetic phase transition from antiferromagnetic to ferromagnetic by regulating spin exchange interactions. This result provides crucial theoretical basis for understanding the origin of the magnetic structure and the control of physical properties in this type of topological magnetic material.

## 4. Conclusion

Polycrystalline $Mn_{3+x}Ga_{1-x}C$ ($0 \leq x \leq 0.60$) samples were synthesized via solid-state reaction. Mn substitution at Ga sites induces lattice contraction and suppresses the low-temperature antiferromagnetic order of $Mn_3GaC$. The canted-to-collinear ferrimagnetic transition temperature drops from 140 K ($x = 0.00$) to 95 K ($x = 0.10$) and vanishes for $x \geq 0.20$, where a robust ferrimagnetic state emerges with a Néel temperature above 400 K. Saturation magnetization peaks at 3.63 $\mu_B$/f.u. ($x = 0.10$) and decreases to 1.52 $\mu_B$/f.u. at $x = 0.60$. Concurrently, the topological Hall resistivity peak shifts to higher temperatures, reaching a maximum of 1.47 $\mu\Omega \cdot cm$ near 300 K at $x = 0.20$. First-principles calculations show that the polar angle of face-centered Mn-I moments relative to [111] reaches ~40° at x = 0.20—consistent with enhanced noncoplanarity and experimental magnetism. With further doping, Mn-I and corner-site Mn-II moments align antiparallel, confirming that inter-sublattice antiferromagnetic coupling drives the magnetic evolution. This work elucidates the exchange mechanism in $Mn_{3+x}Ga_{1-x}C$ and guides the design of high magnetic ordering perovskites.

**Acknowledgments**

The authors gratefully acknowledge the financial support provided by the National Natural Science Foundation of China (Grant No.~51971087), the ``333 Talent Project'' of Hebei Province (Grant No.~C20231105), the Basic Research Project of Shijiazhuang Municipal Universities in Hebei Province (Grant No.~241790617A), the Central Guidance on Local Science and Technology Development Fund of Hebei Province (Grant No.~236Z7606G), and the Science Foundation of Hebei Normal University (Grant No.~L2024B08). These funding sources have been instrumental in facilitating the completion of this research.

kagome lattice: Chiral spin state based on a ferromagnet[J]. Physical Review B, 2000, 62(10): R6065.

[27] Taguchi, Y., Oohara, Y., Yoshizawa, H., et al. Spin Chirality, Berry Phase, and Anomalous Hall Effect in a Frustrated Ferromagnet[J]. Science, 2001, 291: 1573-1576.

[28] Chen, H., Niu, Q., MacDonald, A. H. Anomalous Hall Effect Arising from Noncollinear Antiferromagnetism[J]. Physical Review Letters, 2014, 112: 017205.

[29] Ikeda, T., Tsunoda, M., Oogane, M., et al. Anomalous Hall effect in polycrystalline $Mn_3Sn$ thin films[J]. Applied Physics Letters, 2018, 113: 222405.

[30] Bruno, P., Dugaev, V. K., Taillefumier, M. Topological Hall Effect and Berry Phase in Magnetic Nanostructures[J]. Physical Review Letters, 2004, 93: 096806.

[31] Ikeda, T., Tsunoda, M., Oogane, M., et al. Anomalous Hall effect in polycrystalline $Mn_3Sn$ thin films[J]. Applied Physics Letters, 2018, 113: 222405.

[32] Gong, G. S., Xu, L. M., Bai, Y. M., et al. Large topological Hall effect near room temperature in noncollinear ferromagnet $LaMn_2Ge_2$ single crystal[J]. Physical Review Materials, 2021, 5: 034405.

[33] Yang, S. Y., Wang, Y., Ortiz, B. R., et al. Giant, unconventional anomalous Hall effect in the metallic frustrated magnet candidate, $KV3Sb5$[J]. Science Advances, 2020, 6: eabb6003.